\input ./style/arxiv-general.cfg
\documentclass[aoas,MSNbibl,nameyear,seceqn,dvips]{arximspdf}
\makeatletter
   \@ifpackageloaded{graphicx}{}{\usepackage{graphicx}}
\makeatother
\usepackage{multirow}

\doi{10.1214/15-AOAS814}
\volume{9}
\issue{2}
\pubyear{2015}
\firstpage{883}
\lastpage{900}
\docsubty{FLA}

\makeatletter
\def\sfrac#1#2{#1/#2}
\def\vfrac#1#2{(#1)/#2}

\newcommand{\rrvert}{\vert}
\newcommand{\llvert}{\vert}
\newtheorem{theorem}{Theorem}
\newcommand{\eqref}[1]{(\ref{#1})}
\makeatother

\begin{document}
\begin{frontmatter}

\title{Jump detection in generalized error-in-variables regression with
an application to Australian health tax policies\thanksref{T1}}
\runtitle{Jump detection with error in variables}

\begin{aug}
\author[A]{\fnms{Yicheng}~\snm{Kang}\corref{}\thanksref{M1}\ead[label=e1]{kangx276@umn.edu}},
\author[B]{\fnms{Xiaodong}~\snm{Gong}\thanksref{M2}\ead[label=e2]{Xiaodong.Gong@natsem.canberra.edu.au}},
\author[C]{\fnms{Jiti}~\snm{Gao}\thanksref{M3}\ead[label=e3]{jiti.gao@monash.edu}\ead[label=u3,url]{http://www.buseco.monash.edu.au/about/staff/profile.php?cn=jiti-gao}}
\and
\author[A]{\fnms{Peihua}~\snm{Qiu}\thanksref{M1}\ead[label=e4]{pqiu@ufl.edu}\ead[label=u4,url]{http://users.phhp.ufl.edu/pqiu/}}
\runauthor{Kang, Gong, Gao and Qiu}
\affiliation{University of Florida\thanksmark{M1},
University of Canberra\thanksmark{M2},
Australian National University\thanksmark{M2},
IZA\thanksmark{M2} and
Monash University\thanksmark{M3}}
\address[A]{Y. Kang\\
P. Qiu\\
Department of Biostatistics\\
University of Florida\\
Room 5230\\
2004 Mowry Road\\
Gainesville, Florida 32610\\
USA\\
\printead{e1}\\
\phantom{E-mail: }\printead*{e4}\\
\printead{u4}}
\address[B]{X. Gong\\
NATSEM\\
University of Canberra\\
Canberra, ACT 2601\\
Australia\\
\printead{e2}}
\address[C]{J. Gao\\
Department of Econometrics\\
\quad and Business Statistics\\
Monash University\\
Caulfield East, VIC 3145\\
Australia\\
\printead{e3}\\
\printead{u3}}
\end{aug}
\thankstext{T1}{Supported in part by an NSF grant.}

%
\received{\smonth{10} \syear{2014}}
%
\revised{\smonth{2} \syear{2015}}

%
\begin{abstract}
Without measurement errors in predictors, discontinuity of a nonparametric
regression function at unknown locations could be estimated using a number
of existing approaches. However, it becomes a challenging problem when the
predictors contain measurement errors. In this paper, an error-in-variables
jump point estimator is suggested for a nonparametric generalized
error-in-variables regression model. A major feature of our method is that
it does not impose any parametric distribution on the measurement
error. Its
performance is evaluated by both numerical studies and theoretical
justifications. The method is applied to studying the impact of Medicare
Levy Surcharge on the private health insurance take-up rate in Australia.
\end{abstract}

%
\end{frontmatter}

\setcounter{footnote}{1}

\section{Introduction}
\label{secintroduction}

This paper is motivated by our attempt to study the impact of the
Medical Levy Surcharge (MLS) tax policy on the take-up rate of the
private health insurance (PHI) in Australia. People in Australia
are liable of MLS (which is about 1 percent of their annual taxable
incomes) if they
do not buy PHI and their annual taxable incomes are above a certain
level. For example, the income threshold for single individuals was
\$50,000 per annum in the 2003--2004 financial year, where the dollar
sign ``\$'' used here and throughout the paper represents the Australian
Dollar (AUD). The major purposes of
this tax policy were to give people more choices of health insurance
and to
take a certain pressure off the public medical system. It was expected
that this policy would generate a jump in the PHI take-up rate around the
taxable income threshold. The size of the jump could be used to evaluate
the impact of the policy on the PHI take-up rate. However, the jump
location may not be exactly at the threshold
for the reasons given below. First, the \$500 MLS at the threshold
could be lower than the net cost of PHI, and taxpayers usually consider
buying PHI only when the MLS exceeds the cost of PHI at higher income
levels. Second, the MLS is collected when taxpayers file their tax returns
after the financial year is finished, while the decision to buy the PHI
should be made before the financial year starts. Because it is difficult
for people to predict their taxable incomes accurately, they may not be
aware of the MLS issue until it occurs. So, the jump location is actually
unknown and needs to be estimated properly before the jump size
can be estimated. To estimate the jump location accurately is also
helpful for understanding the demand for PHI in Australia. This is
because from the costs of the PHI and the difference between the
estimated jump point and the threshold, one can infer the true value of PHI
to the taxpayers.

There is a large literature using tax changes as a source of variation in
the after-tax price of health insurances. Most of these
studies are for the US-employer-provided health insurances. See
\citet{GruPot1994}, \citet{Fin2002}, \citet{RodSto2004},
and \citet{BucDidVal2011}
for a few examples. Rarely is it the case that the tax changes could
be argued as exogenous. Jumps caused by policy design, such as the
MLS in Australia, have been argued to be exogenous locally for the individuals
around it [\citet{LeeLem2010}].

In the statistical literature, jump detection in regression functions
has been
discussed by several authors, including \citet{jooqiu2009},
\citeauthor{muller1992} (\citeyear{muller1992,cmuller2002}),
\citeauthor{qiu1991} (\citeyear{qiu1991,qiu1994}),
\citet{qiuyandell1998,wu-chu1993},
and the references therein. See \citet{qiu2005} for an overview on this
topic. All existing jump detection methods assume that the explanatory
variable does not have any measurement error involved.
Meanwhile, the existing literature on the error-in-variables
regression modeling assumes that the measurement error distribution is
known or it can be estimated reasonably well beforehand and that
the related regression function is continuous. See, for example,
\citet
{carroll-et2006,carroll-et1999,comte-taupin2007,cook-stefanski1994,delaigle-meister2007,delaigle2008,fan-masry1992,fan-truong1993,hall-meister2007,liang-wang2005,staudenmayer-ruppert2004,stefanski2000,stefanski-cook1995},
and
\citet{taupin2001}. Our case is much more complicated. The available
data to us are drawn from the ``1\% Sample Unit Record
File of Individual Income Tax Returns'' for the 2003--2004 financial year,
that was developed by the Australian Tax Office (ATO) for research purposes.
Out of privacy consideration, the ATO intentionally perturbed the income
data by multiplying random numbers to the income data, and the true distribution
of the random numbers is unrevealed. Our major task here is to
estimate a jump point and the jump magnitude in a regression model when
the regressor
contains measurement errors with an unknown distribution. This problem
is much more challenging to handle, compared to the ones discussed in
the papers mentioned above. For instance, the deconvolution kernel
regression estimator proposed by \citet{fan-truong1993} assumes that
the characteristic function of the measurement error distribution is
completely known. Such detailed knowledge of the measurement error
distribution is unavailable in the current PHI problem. Also, when the
error distribution was misspecified in the conventional deconvolution
problems, \citet{meister2009} pointed out
that the Mean Integrated Squared Error (MISE) of the deconvolution
kernel estimator
was not bounded from above. Therefore, this estimator can perform badly
when the error distribution is not correctly specified.

In this paper, we propose a generalized error-in-variables regression model
for describing the relationship between the PHI take-up rate and
a person's annual taxable income. In the model, a jump point is
included to
accommodate the possible abrupt impact of MLS on the PHI take-up rate.
A novel jump detector is proposed as well, which takes into
account the measurement errors. One feature of our
method is that it does not require the measurement error distribution to
be specified beforehand, making it applicable to the current PHI problem
and other real problems.

The remainder of the article is organized as follows. In Section~\ref
{secmodell-meth}, our proposed model and jump detector are described in
detail. In Section~\ref{secstat-prop}, some statistical properties
of the proposed jump detector are discussed. In Section~\ref
{secnumerical-studies}, its numerical performance is
evaluated. In Section~\ref{secdata-analysis}, an in-depth analysis of
the PHI data is presented. Several remarks
conclude the article in Section~\ref{secdiscussion}. Some technical details
are provided in a supplementary file.

\section{Proposed methodology}\label{secmodell-meth}

Let $\{(W_i, Y_i)\dvtx  i = 1, \ldots, n \}$ be a sample of $n$ independent and
identically distributed (i.i.d.) observations from the models described
below:
\begin{longlist}[(ii)]
\item[(i)] The conditional
distribution of $Y_i|X_i = x$ has probability density function (p.d.f.)
or probability mass function (p.m.f.) from the exponential family
%
\begin{equation}
\label{eq1} \exp\biggl\{ \frac{y \theta(x) - b(\theta(x))}{a(\phi)} +
c(y, \phi) \biggr\},
\end{equation}
where $X_i$ is the $i$th observation of the unobservable explanatory
variable $X$, $Y_i$ is the $i$th observation of the response variable $Y$,
$\theta(x)$ is the canonical parameter when $X_i=x$, $\phi$ is a scale
parameter, $a(\phi)$, $b(\theta(x))$, and $c(y, \phi)$ are certain
functions of $\phi$, $\theta(x)$, and $(y, \phi)$, respectively.
\item[(ii)] $W_i$ is the observed value of $X_i$ with a measurement error,
and their relationship can be described by the model
%
\begin{equation}
\label{eq2} W_i = X_i + \sigma_n
U_i,
\end{equation}
where $\sigma_n>0$ denotes the standard deviation of the measurement error
in $X_i$, and $U_i$ is the standardized measurement error with mean 0
and variance
1. It is also assumed that $U_i$'s are i.i.d., $U_i$ is independent of
both $X_i$ and $Y_i$, the distribution of $U_i$,
denoted as $f_U$, and the distribution of $X_i$, denoted as $f_X$, are both
unknown.
\end{longlist}

In model \eqref{eq1}, $\theta(x)$ relates $Y_i$ to $X_i$. And, model
\eqref{eq1} includes many
commonly used distributions (e.g., Normal, Poisson, Binomial) as special
cases. In the current PHI problem, because ATO perturbed the income data
by multiplying each original income observation by a random number, the income
variables are used in $\log$ scale so that model \eqref{eq2} with
additive measurement error is appropriate. More specifically, $X_i$ and
$W_i$ denote the true and observed annual taxable incomes in $\log$
scale, respectively, $Y_i$ denotes the status of
PHI take-up ($Y_i$ equals~1 when a specific person buys the PHI and 0
otherwise), and $Y_i|X_i=x$ is assumed to follow the Bernoulli
distribution with the probability of success being
$p(x)=P(Y_i=1|X_i=x)$. In such cases, the quantities $\theta(x)$,
$a(\phi)$, $b(\theta(x))$, and $c(y, \phi)$ can be specified as
follows:
\begin{eqnarray*}
\theta(x) &=& \log\biggl(\frac{p(x)}{1 - p(x)} \biggr), \qquad  a(\phi) = 1,
\\
b(\theta) &=& \log\bigl(1 + \exp(\theta)\bigr),\qquad  c(y, \phi) = 0.
\end{eqnarray*}
As discussed in Section~\ref{secintroduction}, the tax policy MLS is expected
to generate a jump in $p(x)$. So, a jump in $\theta(x)$ is expected as well.

\begin{figure}

\includegraphics{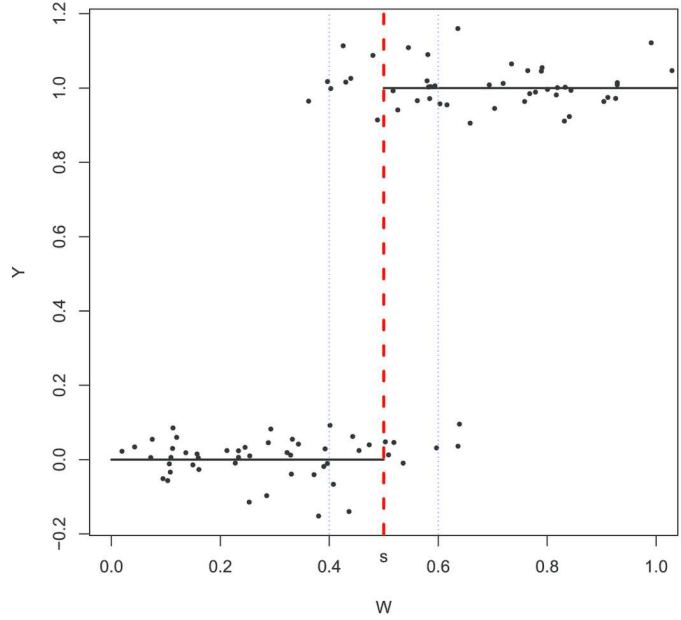}

\caption{The solid line denotes the conditional mean function $m(x)=
E(Y|X=x)$ that has a jump at $s=0.5$ (marked by the vertical dashed
line). The dark points denote
observations of $(W, Y)$ where $W$ is the observed value of
$X$ with measurement error involved. It can be seen that the
values of $X$ corresponding to those $W$ values that
are close to the true jump location (e.g., those fall between
the two vertical dotted lines) are likely to be on both sides
of the jump.}
\label{fig1}
\end{figure}

In the PHI problem, it is important to estimate the jump position in
$\theta(x)$
in order to study the impact of the tax policy MLS on the PHI take-up. Our
proposed jump detector is described below. The true jump
position of $\theta(x)$ is assumed to be at $s$ which is unknown.
Without loss
of generality, let us assume that the support of $f_X$ is $[0, 1]$ and
$s \in(0, 1)$. Let $m(x)$ denote the conditional mean of $Y$ given $X=x$.
Then, it can be checked from \eqref{eq1} that
\[
m(x) = E(Y|X=x) = b'\bigl(\theta(x)\bigr).
\]
For any given point $x \in(2 h_n, 1 - 2 h_n)$,
let us consider its right-sided neighborhood $[x, x + h_n]$, where $h_n
> 0$
is a bandwidth parameter. When there is no measurement error in $X$,
$m(x+)=\lim_{\Delta x \rightarrow0+} m(x + \Delta x)$ can be estimated
reasonably well [cf. \citet{qiu2005}, Chapter~2], by
%
\begin{equation}
\label{eq3} \sum_{i=1}^n
Y_i K_r \biggl( \frac{X_i - x}{h_n} \biggr)\Big/ \sum
_{i=1}^n K_r \biggl(
\frac{X_i - x}{h_n} \biggr),
\end{equation}
where $K_r$ is a decreasing kernel function with the right-sided
support $(0, 1]$.
In the case when $X$ has measurement errors involved, the estimator in
\eqref{eq3} is unavailable because $X_i$'s are no longer observable.
It may be problematic if we simply replace $X_i$'s by $W_i$'s in \eqref{eq3}
because we do not know whether a specific $X_i$ is located on the right-hand side
of $x$ or not when its observed value $W_i$ is on the right-hand side of
$x$, due
to the measurement error. However,
as demonstrated in Figure~\ref{fig1}, the following fact can be observed:
if $W_i$ is close to the true jump point $s$, then the corresponding
unobservable $X_i$ is likely to be on the other side of the jump location
and, consequently, $Y_i$ follows a distribution with
the parameter $\theta(X_i)$ which could be very different from
$\theta(W_i)$. A one-sided kernel estimator defined in \eqref{eq3}
with the $X_i$'s replaced by the corresponding $W_i$'s
actually averages observations on both sides of the jump
location. Thus, the impact of the measurement error could be severe
in such cases. On the other hand, in the
case when $W_i$ is farther away from $s$, such an impact becomes smaller.
%
Based on this fact, let us consider a one-step-right neighborhood of $x$,
defined to be $N_{n,r}(x;h_n):= (x + h_n, x + 2h_n)$, and define
%
\begin{equation}
\label{eq4} \qquad\widehat{m}_{n,r}(x+) = \sum
_{i = 1}^n Y_i K_r \biggl(
\frac{W_i
- (x + h_n)}{h_n} \biggr) \Big/ \sum_{i = 1}^n
K_r \biggl( \frac{W_i
- (x + h_n)}{h_n} \biggr),
\end{equation}
where the kernel function $K_r$ is the same as the one in \eqref{eq3}.
If $x$ is at the jump location, then $\widehat{m}_{n,r}(x+)$ should be
a weighted average of observations that are on the right-hand side of
$x$ since the observations in the neighborhood $N_{n,r}(x;h_n)$ are all
quite far away from the
jump point. This would limit the possibility of averaging observations on
both sides of $x$, and thus diminish the impact of the measurement error.
However, this estimator may still have some bias for estimating $m(x+)$
because the $X$ values of most observations used in $\widehat{m}_{n,r}(x+)$
are at least one bandwidth above $x$. To address this issue, let us
define
%
\begin{equation}
\label{eq5} \qquad\widehat{m}_n(x+) = \frac{\sum_{i = 1}^n Y_i K_r (
\vfrac{W_i - x}{h_n} ) K^* ( \sfrac{|\widehat{m}_n^*(W_i+) -
\widehat{m}_{n,r}(x+)|}{\rho_n} )} {
\sum_{i = 1}^n K_r (
\vfrac{W_i - x}{h_n} ) K^* ( \sfrac{|\widehat{m}_n^*(W_i+) -
\widehat{m}_{n,r}(x+)|}{\rho_n} )},
\end{equation}
where the kernel function $K_r$ is the same as the one in
\eqref{eq3}, $K^*$ is another decreasing kernel function with support
$[0, 1]$,
$\rho_n = \max_{x \leq W_i \leq x + h_n} \allowbreak
|\widehat{m}_n^*(W_i+) - \widehat{m}_{n,r}(x+)|$, and
\[
\widehat{m}_n^*(x+) = \sum_{j=1}^n
Y_j K_r \biggl( \frac{W_j -
x}{h_n} \biggr) \Big/ \sum
_{j=1}^n K_r \biggl(
\frac{W_j
-x}{h_n} \biggr)
\]
is the conventional one-sided kernel estimator of $m(x+)$. The intuitive
explanation of \eqref{eq5} is as follows. From its
definition, $\widehat{m}_n^*(W_i+)$ is mainly determined by
observations close to $W_i$ and these observations are mostly in the
neighborhood $[x, x+h_n]$. For some of these observations, the
corresponding $X_i$'s may be on the left-hand side of $x$ and, thus, the
impact of the measurement error on the one-sided estimator in
\eqref{eq5} could be severe. On the other hand, the measurement
error does not have much impact on $\widehat{m}_{n,r}(x+)$,
as explained above. The bias in $\widehat{m}_{n,r}(x+)$ for estimating
$m(x+)$, due to the fact that it uses many observations that are
at least one bandwidth on the right-hand side of $x$, is considered
significantly smaller than the bias in $\widehat{m}_n^*(W_i+)$ due to
measurement error, because the former is due to the variation of
$m(\cdot)$ in a small continuity region while the latter is caused
by the jump. So, the difference
$\widehat{m}_n^*(W_i+)-\widehat{m}_{n,r}(x+)$ can provide us a measure
of the impact of the measurement error in $W_i$ on estimation of $m(x+)$.
If the difference is small, then the impact of the measurement error in
$W_i$ should be small. Otherwise, its impact should be large. The
kernel function $K^*$ in \eqref{eq5} aims to eliminate such an impact.
Therefore,
$\widehat{m}_n(x+)$ should provide a reasonable estimator for $m(x+)$.
An estimator of $m(x-)$ can be constructed similarly to \eqref{eq5},
which is denoted as $\widehat{m}_n(x-)$. Then, the true jump location $s$
can be estimated by
%
\begin{equation}
\label{eq7} \widehat{s}_n = \operatorname{arg}\max
_{x \in(2 h_n, 1 - 2 h_n)} \bigl\llvert\widehat{m}_n(x+) -
\widehat{m}_n(x-) \bigr\rrvert,
\end{equation}
and the corresponding jump magnitude $d$ in $m(x)$ can be estimated by
%
\begin{equation}
\widehat{d}_n = \widehat{m}_n(\widehat{s}_n+)-
\widehat{m}_n(\widehat{s}_n-). \label{eqjs}
\end{equation}
It should be pointed out that, although the exponential family in
\eqref{eq1} is parameterized using the canonical parameter $\theta(x)$,
the mean parameter $m(x)$ is often easier to interpret in practice.
For this reason, both the jump location and the jump magnitude are
discussed above in terms of $m(x)$, instead of $\theta(x)$.
In Section~\ref{secstat-prop}, we will show that under certain
regularity conditions, $\widehat{s}_n$ is a consistent estimator.

In the proposed jump detector \eqref{eq7}, there is one parameter $h_n$
to choose. According to
\citet{gijbels-goderniaux2004}, jump detectors based on kernel smoothing
in cases without measurement error depend heavily on the choice of
bandwidth parameters. In simulation studies, the true jump location
$s$ could be
known. Then, $h_n$ can be chosen to be the one that minimizes
$|\widehat{s}_n(h_n) - s|$, where $\widehat{s}_n$ has been written as
$\widehat{s}_n(h_n)$ for convenience of discussion. In practice, $s$
is usually unknown. In such cases,
we suggest the following bootstrap bandwidth selection procedure:
\begin{itemize}
\item For a given bandwidth value $h_n > 0$, apply the proposed jump detection
procedure \eqref{eq4}--\eqref{eq7} to the original data set $\{(W_1, Y_1)$,
$(W_2, Y_2), \ldots,\break (W_n, Y_n)\}$, and obtain an estimator of $s$,
denoted as $\widehat{s}_n(h_n)$.
\item Draw with replacement $n$ times from the original data set to
obtain\break the first bootstrap sample, denoted as
$\{(\widetilde{W}{}^{(1)}_1, \widetilde{Y}{}^{(1)}_1)$, $
(\widetilde{W}{}^{(1)}_2, \widetilde{Y}{}^{(1)}_2), \ldots,\break
(\widetilde{W}{}^{(1)}_n, \widetilde{Y}{}^{(1)}_n)\}$.
\item Apply the proposed jump detection procedure
\eqref{eq4}--\eqref{eq7} to the first bootstrap sample, and obtain
the first bootstrap estimator of $s$, denoted as $\widetilde{s}_n^{(1)}(h_n)$.
\item Repeat\vspace*{1pt} the previous two steps $B$ times and obtain $B$ bootstrap
estimators of $s$:
$\{\widetilde{s}_n^{(1)}(h_n), \widetilde{s}_n^{(2)}(h_n), \ldots,
\widetilde{s}_n^{(B)}(h_n)\}$.
\item Then, the bandwidth $h_n$ is chosen to be the minimizer of
%
\begin{equation}
\label{eq8} \min_{h_n} \frac{1}{B} \sum
_{k = 1}^B \bigl\llvert\widehat{s}_n(h_n)
- \widetilde{s}_n^{(k)}(h_n)\bigr\rrvert.
\end{equation}
\end{itemize}

It should be pointed out that, as a byproduct of the above bootstrap
bandwidth selection procedure, a confidence interval for $s$
can be constructed from the empirical distribution of
$\{\widetilde{s}_n^{(1)}(\widetilde{h}_n),
\widetilde{s}_n^{(2)}(\widetilde{h}_n), \ldots,
\widetilde{s}_n^{(B)}(\widetilde{h}_n)\}$, where $\widetilde{h}_n$
denotes the bandwidth selected by the bootstrap. More specifically, for a
given significance level $\alpha\in(0, 1)$, a $100(1-\alpha)\%$ confidence
interval for $s$ is defined to be $ (\widetilde{s}_{n,\alpha
/2}(\widetilde{h}_n),
\widetilde{s}_{n,1 - \alpha/2}(\widetilde{h}_n) )$, where
$\widetilde{s}_{n,\alpha/2}(\widetilde{h}_n)$ and
$\widetilde{s}_{n,1- \alpha/2}(\widetilde{h}_n) $ denote the $(\alpha/2)$th
and $(1 -\alpha/2)$th quantiles of the empirical distribution of
$\{\widetilde{s}_n^{(1)}(\widetilde{h}_n),
\widetilde{s}_n^{(2)}(\widetilde{h}_n), \ldots,
\widetilde{s}_n^{(B)}(\widetilde{h}_n)\}$.

\section{Statistical properties}\label{secstat-prop}

In this section, we discuss some statistical properties of the proposed jump
detector defined in \eqref{eq7}. To this end, we have the theorem
below.

\begin{theorem}
\label{thm1}
Assume that $\{(W_1, Y_1), (W_2, Y_2), \ldots, (W_n, Y_n)\}$ are
i.i.d. observations from models \eqref{eq1} and
\eqref{eq2}, and the following conditions are satisfied:
\begin{longlist}[(1)]
\item[(1)] $\theta(\cdot)$ is a bounded, piecewise continuous function
with a single jump at $s$ and its first-order derivative is also a
bounded function,
\item[(2)] $a(\cdot)$, $b(\cdot)$, $b'(\cdot)$, and $b''(\cdot)$ are all
bounded and continuous functions,
\item[(3)] $(b')^{-1}(\cdot)$ exists and it is strictly monotone and
Lipschitz-1 continuous\footnote{Given an interval $I \subset
\mathcal{R}$, a function $g: I \rightarrow
\mathcal{R}$ is called Lipschitz-1 continuous if there exists a
real constant $C\geq0$ such that, for all $x_1$ and $x_2 \in
I$, $|g(x_1) - g(x_2)| \leq C|x_1 - x_2|$.} in any compact
subset of the range of $\theta(\cdot)$,
\item[(4)] the support of $f_X$ is $[0,1]$ and $s \in(0, 1)$,
\item[(5)] $f_X$ is continuous, bounded, and positive on $(0,1)$,
\item[(6)] $f_U$ is a continuous function and has a positive value at 0,
\item[(7)] the kernel functions $K^*$ and $K_r$ are Lipschitz-1
continuous density functions with the same support $[0, 1]$,
\item[(8)] the\vspace*{1pt} bandwidth $h_n$ satisfies the conditions that $h_n = o(1)$,
and $(\log
n)^{1 + \eta} /\break  (n^{1/2}h_n) = o(1)$, for some $\eta> 1/2$.
\end{longlist}
Then, we
have the following results:
\begin{longlist}[(ii)]
\item[(i)] If $\sigma_n / h_n = o(1)$, then
\[
\widehat{m}_n(x+) - \widehat{m}_n(x-) = m(x+) - m(x-) +
O \biggl(\frac{\sigma_n ^{2/3}}{h_n ^{2/3}} \biggr) + O ( \beta_n
)\qquad\mbox{a.s.},
\]
where $\beta_n = h_n + \frac{(\log n)^{1 + \eta}}{n^{1/2}h_n}$.
\item[(ii)] If $\lim_{n \rightarrow\infty}\sigma_n / h_n = C $, for some $C
> 0$, then
\begin{eqnarray*}
&& \widehat{m}_n(x+) - \widehat{m}_n(x-)
\\
&&\qquad  =  m(x+) -
m(x-) - \bigl(m(x+) - m(x-) \bigr) C_{K,r}
+ O(\beta_n)\qquad\mbox{a.s.,}
\end{eqnarray*}
where
\[
C_{K,r} = \frac{\int_0^1 K_r^{**}(w) P(\llvert U\rrvert > w/C) \,dw}{\int_0^1
K_r^{**}(w) \,dw},
\]
and
\[
K_r^{**}(w) = K_r(w) K^* \biggl(
\frac{\int_0^1 K_r(v)P(v
+ w < CU < v + 1) \,dv}{\int_0^1 K_r(v)P(v < CU < v + 1) \,dv} \biggr).
\]
\item[(iii)] If the conditions in either \textup{(i)} or \textup{(ii)} hold, then we have
\[
\llvert\widehat{s}_n - s \rrvert= O(h_n), \qquad a.s.
\]
\end{longlist}
\end{theorem}

Theorem \ref{thm1} shows that the jump detector defined in \eqref{eq7}
provides a statistically consistent estimator of $s$ under some regularity
conditions. Its proof is given in a supplementary file [\citet
{kang-et2015}]. In result (ii)
of the
above theorem, if $K^*$ and $K_r$ are both decreasing on $[0, 1]$,
then we have
\[
\frac{\int_0^1 K_r^{**}(w) P(\llvert U\rrvert > w/C) \,dw}{\int_0^1
K_r^{**}(w) \,dw} \leq\frac{\int_0^1 K_r(w) P(\llvert U\rrvert > w/C)
\,dw}{\int_0^1 K_r(w) \,dw}.
\]
It can be checked that if the conventional kernel estimators are used
when defining the jump detection criterion [i.e., $\widehat{m}_n^*(x+)$ is
used], then the asymptotic bias for $\widehat{m}_n^*(x+) - \widehat
{m}_n^*(x-)$ to
estimate $m(x+) - m(x-)$ is
$(m(x+) - m(x-))\frac{\int_0^1 K_r(w) P(|U| > w/C) \,dw}{
\int_0^1 K_r(w) \,dw}$, which is larger than the asymptotic bias\break
$(m(x+) - m(x-)) \frac{\int_0^1 K_r^{**}(w) P(|U| > w/C) \,dw}{
\int_0^1 K_r^{**}(w) \,dw}$ when we use $\widehat{m}_n(x+) - \widehat{m}_n(x-)$
to estimate $m(x+) - m(x-)$. Therefore, the second kernel function $K^*$
used in \eqref{eq5} is helpful in reducing the asymptotic bias. In
Theorem \ref{thm1},
it is required that the measurement error variance $\sigma_n^2$ tends
to 0
when the sample size $n$ increases. In the literature, it has been pointed
out that this condition is needed for consistently estimating the regression
function when its observations have measurement errors involved and when
little prior information about the measurement error distribution is
available
[cf. \citet{delaigle2008}].

\section{Numerical studies}
\label{secnumerical-studies}

In this section, we present some results regarding the
numerical performance of the proposed jump detector described in
Section~\ref{secmodell-meth}, which are organized in two
subsections. Section~\ref{secnumer-perf-prop} includes some
simulation examples related to the jump detector defined in
\eqref{eq7}. Section~\ref{seccomp-naive-estim} compares the
proposed jump detector to the difference-kernel-estimation
(DKE) procedure that ignores the measurement error
[cf. \citet{qiu2005}, Section~3.2].

\subsection{Numerical performance of the proposed jump detector}\label{secnumer-perf-prop}

In this subsection, the performance of the proposed jump detector is evaluated
using two simulated examples. In each example, we consider cases when the
sample size $n$ equals 100 or 200, $f_X \sim \operatorname{Unif}[0, 1]$, and $f_U$
is either a Normal, a Laplace, or a Uniform distribution, with $E(U) =
0$ and
$\operatorname{Var}(U)/\operatorname{Var}(X)$ fixed at 15\%. In each combination of $n$ and
$f_U$, the simulation is repeated 100 times. For each given bandwidth $h_n$,
100 values of the Absolute Error (AE), defined as AE$(h_n)=|\widehat{s}_n(h_n)
- s|$, are computed. Their average is called the Mean Absolute Error (MAE)
and is denoted as $\operatorname{MAE}(h_n)$. The minimizer of $\operatorname{MAE}(h_n)$ is called the
\textit{optimal bandwidth} and is denoted as $h_{\mathrm{opt}}$. In each replicated
simulation, we also compute a bandwidth value using the proposed bootstrap
procedure. The average of such 100 bandwidth values is called the
\textit{bootstrap bandwidth}, denoted as $h_{\mathrm{bt}}$. In each example,\vadjust{\goodbreak} the
values of $h_{\mathrm{opt}}$, $h_{\mathrm{bt}}$, $\operatorname{MAE}(h_{\mathrm{opt}})$, $\operatorname{MAE}(h_{\mathrm{bt}})$, and the
empirical coverage probability (CP) of the $95\%$ confidence interval (see
its description at the end of Section~\ref{secmodell-meth}) computed
from the 100 replicated simulations are presented. In the case when
$n = 100$ and $f_U$ is Normal, the sample that gives the median value of
AE$(h_{\mathrm{opt}})$ is denoted as $\mathcal{S}_{50}$. For that sample, the
estimated jump location by \eqref{eq7} with $h_n=h_{\mathrm{bt}}$ and the
corresponding $95\%$ confidence interval for $s$ will be presented.
Throughout this section, if there is no further specification, the
bootstrap sample
size $B$ is chosen to be 999, and $K^*$ and $K_r$ used in
\eqref{eq4}--\eqref{eq7} are both chosen to be the Epanechnikov kernel
function.

%
\begin{figure}[b]

\includegraphics{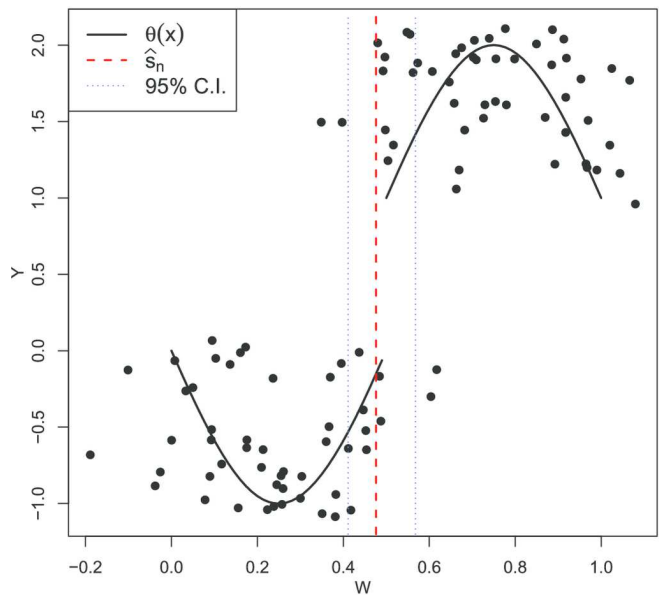}

\caption{A realization of $\mathcal{S}_{50}$ in the first example with
the true function of $\theta(\cdot)$ (solid line), the estimated jump location
$\widehat{s}_n$ (vertical dashed line), and a $95\%$ confidence interval
for the true jump location $s$ (vertical dotted lines).}
\label{fig2}
\end{figure}

In the first example, the conditional distribution of $Y|X=x$ is
assumed to
follow the Normal distribution with density
\[
\frac{1}{\sqrt{2\pi\times0.01^2}} \exp\biggl\{ - \frac{(y - \theta
(x))^2}{2 \times0.01^2} \biggr\},
\]
where
\[
\theta(x) = \cases{ -\sin(2\pi x), &\quad if $x \leq0.5$;
\cr
-\sin(2\pi x) +
1, &\quad otherwise.}
\]
Figure~\ref{fig2} shows a realization of the sample $\mathcal{S}_{50}$,
the true function $\theta(\cdot)$ (solid line), the estimated jump location
$\widehat{s}_n$ (vertical dashed line), and a $95\%$ confidence interval
for the true jump location $s$ (vertical dotted lines). It can be seen
that the
proposed jump detector estimates the true jump location reasonably well.
The numerical performance of the jump detector \eqref{eq7} based on
100 replicated simulations is summarized in Table~\ref{tab1}. From the
table, it can be seen that (i) the proposed jump
detector estimates the true jump location reasonably well for various
error distributions, (ii) the performance of $\widehat{s}_n$ improves
as the sample size
$n$ increases, (iii)~the bootstrap bandwidths are slightly larger than the
optimal bandwidths but they are quite close to each other, and
(iv) the empirical coverage probabilities of the proposed confidence
interval for $s$ are all close to the nominal coverage probability~0.95.

%
\begin{table}[t]
\tabcolsep=0pt
\caption{Numerical summary of the first simulation example based on 100 replicated simulations}\label{tab1}
\begin{tabular*}{\tablewidth}{@{\extracolsep{\fill}}lcccccc@{}}
\hline
$\bolds{n}$ & $\bolds{f_U}$ & $\bolds{h_{\mathrm{opt}}}$ & $\bolds{h_{\mathrm{bt}}}$ & $\mathbf{MAE}\bolds{(h_{\mathrm{opt}})}$
& $\mathbf{MAE}\bolds{(h_{\mathrm{bt}})}$ & \textbf{CP}\\
\hline
{100} & Normal &0.3000 &0.3008 &0.0290 &0.0293 &0.95 \\
& Laplace &0.2931 &0.2985 &0.0292 &0.0308 &0.98 \\
& Uniform &0.2767 &0.2910 &0.0335 &0.0347 &0.96
\\[3pt]
200 & Normal &0.2991 &0.3074 &0.0232 &0.0245 &0.96 \\
& Laplace &0.2902 &0.3002 &0.0191 &0.0195 &0.94 \\
& Uniform &0.2721 &0.2917 &0.0232 &0.0239 &0.97 \\
\hline
\end{tabular*}
\end{table}

\begin{figure}

\includegraphics{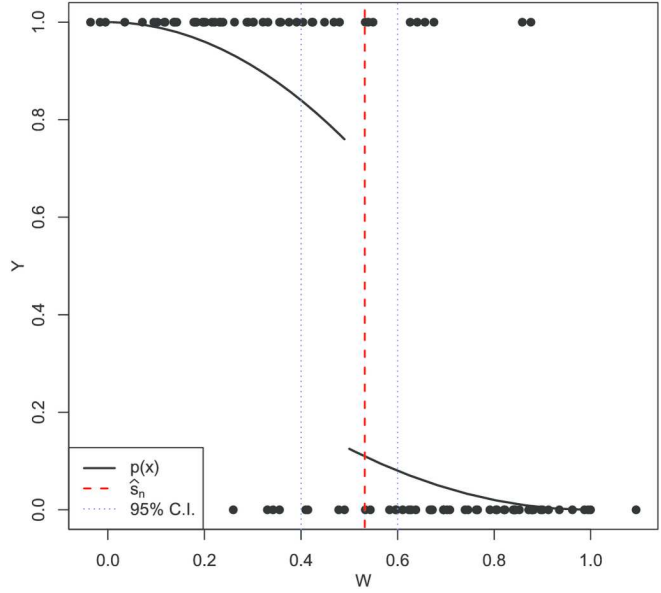}

\caption{A realization of
$\mathcal{S}_{50}$ in the second example with the true function of
$p(\cdot)$ (solid line), the estimated jump location $\widehat{s}_n$
(vertical dashed line), and the $95\%$ confidence interval for $s$
(vertical dotted lines).}
\label{fig3}
\end{figure}

Next, we discuss the second simulation example whose setting is made
similar to that of the PHI data. Assume that the conditional
distribution of
$Y|X=x$ is Bernoulli with the probability of success being
\[
p(x) = \cases{ 1 - x^2, &\quad if $x \in(0, 0.5]$;
\cr
0.5(1 -
x)^2, &\quad if $x \in(0.5, 1)$.}
\]
Figure~\ref{fig3} shows a realization of $\mathcal{S}_{50}$ with
the true function of $p(\cdot)$ (solid line), the estimated jump location
$\widehat{s}_n$ (vertical dashed line), and the $95\%$ confidence
interval for $s$ (vertical dotted lines). It can be seen from the figure
that the sample $\mathcal{S}_{50}$ has quite severe measurement errors
involved and
that the proposed jump detector gives a reasonably good estimate of
the true jump location. The numerical performance of the jump detector
\eqref{eq7} based on 100 replicated simulations is summarized in
Table~\ref{tab2}. From the table, it can be seen that similar
conclusions to those
in the first example can be made here.
%
%
%
\begin{table}[b]
\tabcolsep=0pt
\caption{Numerical summary of the second simulation example based on 100 replicated simulations}\label{tab2}
\begin{tabular*}{\tablewidth}{@{\extracolsep{\fill}}lcccccc@{}}
\hline
$\bolds{n}$ & $\bolds{f_U}$ & $\bolds{h_{\mathrm{opt}}}$ & $\bolds{h_{\mathrm{bt}}}$ & $\mathbf{MAE}\bolds{(h_{\mathrm{opt}})}$
& $\mathbf{MAE}\bolds{(h_{\mathrm{bt}})}$ & \textbf{CP}\\
\hline
{100} & Normal &0.3329 &0.3448 &0.0407 &0.0444 &0.93 \\
& Laplace &0.2758 &0.2826 &0.0397 &0.0422 &0.98\\
& Uniform &0.3203 &0.3247 &0.0479 &0.0484 &0.97
\\[3pt]
{200} & Normal &0.3122 &0.3100 &0.0363 &0.0369 &0.98 \\
& Laplace &0.2820 &0.2816 &0.0326 &0.0336 &0.92\\
& Uniform &0.2878 &0.2983 &0.0352 &0.0352 &0.94 \\
\hline
\end{tabular*}
\end{table}

%
%
\begin{figure}[t]

\includegraphics{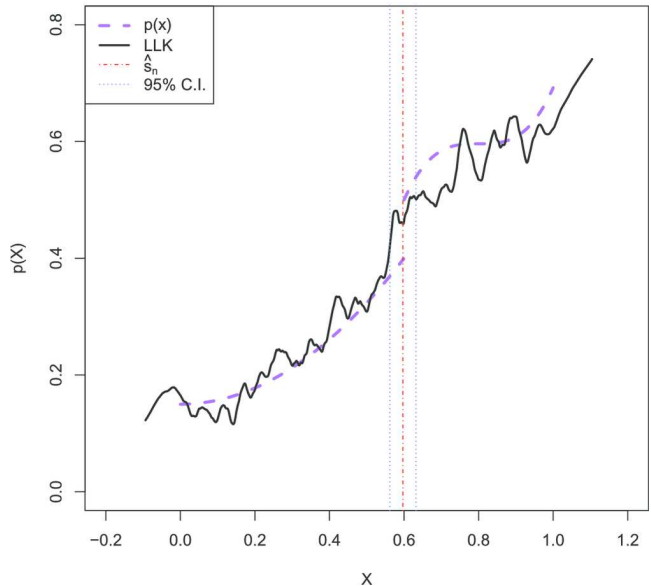}

\caption{The dashed line denotes the true curve of $p(\cdot)$,
the solid line denotes the local linear kernel (LLK)
estimate of $p(\cdot)$ using one realization of simulated data
when $f_X(\cdot)$ is $\operatorname{Unif}[0, 1]$, the vertical dot-dashed line
denotes the estimated jump location $\widehat{s}_n$, and the
vertical dotted lines denote the $95\%$ confidence interval for $s$.}
\label{figcompare}
\end{figure}

%
\begin{figure}[t]

\includegraphics{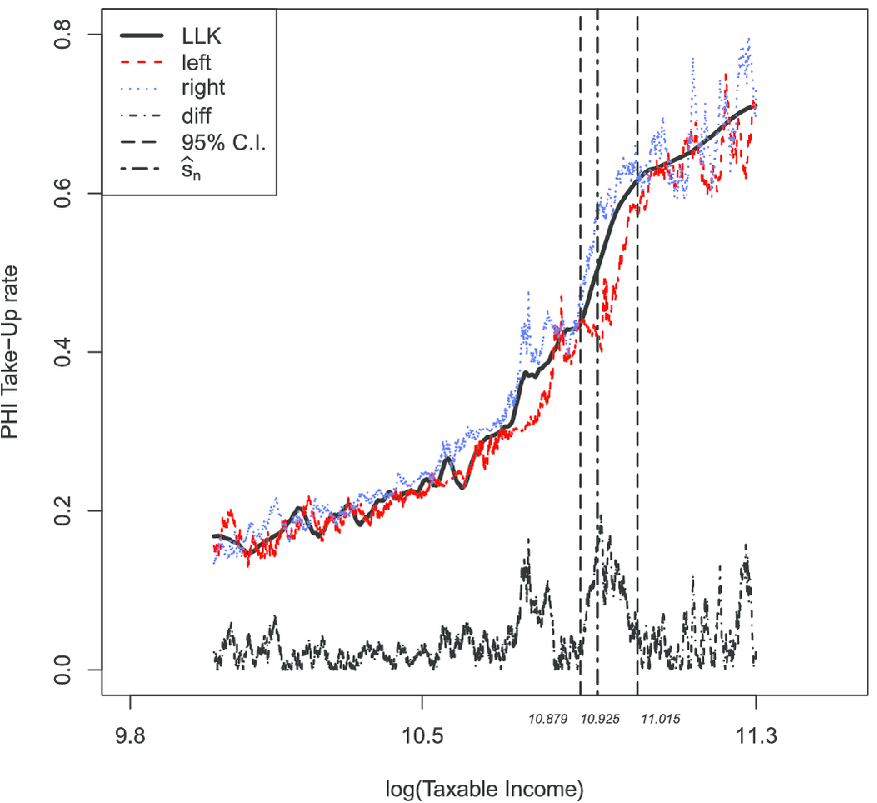}

\caption{The solid line denotes the local linear kernel (LLK)
estimate of $p(\cdot)$ in the PHI example, the dashed line denotes
the left-sided estimate of $p(\cdot)$, the dotted line denotes the
right-sided estimate of $p(\cdot)$, the dot-dash line denotes
the absolute difference between the two
one-sided estimates of $p(\cdot)$, the long-dash vertical lines
denote the 95\% confidence
interval for $s$, and the two-dash line denotes the estimated jump
location $\widehat{s}_n$.}\vspace*{-3pt}
\label{fig4}
\end{figure}

\subsection{Comparison to the DKE estimator}\label{seccomp-naive-estim}
The DKE procedure [see
Section~3.2 in \citet{qiu2005} for a detailed discussion]
provides a
good estimator of the true jump position when there is no measurement
error involved. In this subsection, we compare our proposed jump
detector \eqref{eq7} with the DKE procedure in an artificial
example with similar setup to that of the PHI data. The proposed jump
detector is denoted as NEW and the DKE procedure is denoted as
DKE. Assume that the conditional distribution of $Y|X=x$ is
Bernoulli with the probability of success being
\[
p(x) = \cases{ \frac{25}{36}x^2 + 0.15, &\quad if $x \in[0,
0.6)$;
\vspace*{3pt}\cr
12(x - 0.8)^3 + 0.596, &\quad if $x \in[0.6, 1]$.}
\]
It can be seen that $p(x)$ is a piecewise polynomial with a jump
of size $0.1$ at $x=0.6$, as plotted in Figure~\ref{figcompare}. In
the figure, the estimated function of $p(\cdot)$ by the
local linear kernel (LLK) smoothing procedure is also shown. From the
plot, it can be seen that the shape of $p(\cdot)$ is similar to that
in the PHI study, which is shown in Figure~\ref{fig4}. Note that the
jump size in the PHI application is estimated to be $0.19$ (cf.
Section~\ref{secdata-analysis}), while the jump size in this example is about
half of that size. To make this numerical study in a similar setup to that
of the PHI study, which has 9685 observations distributed in the interval
$[9.8, 11.3]$, we choose the sample size $n$ in the current example to be
6000 and the observations are in the design interval $[0, 1]$. In addition,
$f_U$ is chosen to be $N(0, 0.05^2)$, and $f_X$ is either $\operatorname{Unif}[0, 1]$,
$\operatorname{Beta}(2, 2)$, $\operatorname{Beta}(3, 2)$, or $\operatorname{Beta}(2,3)$. In each case, the simulation
is repeated 100 times, the optimal bandwidth is selected based on the 100
replicated simulations, and the mean and standard deviation of the 100 values
of $|\widehat{s}_n -s|$ and the 100 values of $|(\widehat{m}_n(\widehat
{s}_n+) -
\widehat{m}_n(\widehat{s}_n-))-0.1|$ (i.e., the absolute bias of the
estimated jump size) are computed, respectively. They are denoted as MAE,
SDAE, MABJS, and SDABJS. The results are presented in Table~\ref{tab3}.

%
%
\begin{table}[b]
\tabcolsep=0pt
\caption{Numerical comparison of the proposed jump detector NEW
with the DKE procedure based on 100 replicated simulations. MAE and
SDAE denote the mean and standard deviation of the 100 values of
$|\widehat{s}_n -s|$. MABJS and SDABJS denote the mean and standard
deviation of the 100 values of $|(\widehat{m}_n(\widehat{s}_n+) -
\widehat{m}_n(\widehat{s}_n-))-0.1|$}\label{tab3}
\begin{tabular*}{\tablewidth}{@{\extracolsep{\fill}}@{}lcccccccc@{}}
\hline
& \multicolumn{4}{c}{\textbf{DKE}} & \multicolumn{4}{c@{}}{\textbf{NEW}}\\[-6pt]
& \multicolumn{4}{c}{\hrulefill} & \multicolumn{4}{c@{}}{\hrulefill}\\
$\bolds{f_X}$ & \textbf{MAE} & \textbf{SDAE} & \textbf{MABJS} & \textbf{SDABJS} & \textbf{MAE} & \textbf{SDAE} & \textbf{MABJS} & \textbf{SDABJS} \\
\hline
$\operatorname{Unif}[0,1]$ &0.01718 &0.00126 &0.02598 &0.00064 &0.01532 &0.00115 &0.00511&0.00047 \\
$\operatorname{Beta}(2,2)$ &0.01547 &0.00109 &0.02475 &0.00049 &0.01329 &0.00108 &0.00961 &0.00044 \\
$\operatorname{Beta}(3,2)$ &0.01607 &0.00110 &0.02593 &0.00040 &0.01305 &0.00092 &0.00494&0.00037 \\
$\operatorname{Beta}(2,3)$ &0.01810 &0.00113 &0.02707 &0.00055 &0.01690 &0.00108 &0.01031&0.00058 \\
\hline
\end{tabular*}\vspace*{-3pt}
\end{table}

\noindent

From Table~\ref{tab3}, it can be seen that when there is measurement
error involved, (i)~the precision of the detected jump by the proposed
jump detection procedure is better than that of the DKE
procedure, across all different choices of $f_X$, (ii) the
proposed procedure reduces the bias of the estimated jump size, which is
consistent with our discussion in Section~\ref{secstat-prop}, and (iii)
the proposed procedure has a slightly smaller variability for detecting the
jump location and about the same variability for estimating the jump
size, compared to the DKE procedure.

\section{Analysis of the PHI data}\label{secdata-analysis}

In this section, we apply our proposed jump detector to the PHI
data for evaluating the impact of MLS on the take-up rate of~PHI.\vadjust{\goodbreak}

The purposes of introducing PHI in Australia
were to give consumers more choices and take some pressure off the
public medical system. However, the PHI take-up rate by Australians
was very low at the beginning when the PHI was first introduced in 1984,
and the take-up rate had been declining toward the end of 1990s
(the take-up rate was only about 31 percent at that time) until a
series of policies (including MLS) were introduced.
Impact of some of these policy measures (e.g., Lifetime Health Cover)
have been studied in a few studies, including \citet{butler2002},
\citet{frech-et2003}, \citet{palan-yong2005}, and \citet
{palan-et2009}.
But the role of MLS has not been identified separately yet. The MLS
was imposed in
1997 on high-income taxpayers who did not have private insurances.
Between 1997--1998 and 2007--2008, the threshold of annual taxable
income at which MLS was payable was \$50,000 for singles without
children and \$100,000 for couples. For each dependent child in the
household, the threshold increased by \$3000.
So, people having children may lead
to multiple jumps in the current PHI data. Unfortunately, we do not
have information on the number of children in a family. Also, multiple
jump locations within a relatively narrow range would be difficult
to distinguish, given the measurement error involved in the PHI
data. To mitigate the effect of multiple
jumps due to people having children, this paper only focuses on
singles in the current PHI data.

The data used here are from a confidentialized ``1\% Sample Unit Record
File of Individual Income Tax Returns'' for the 2003--2004 financial year,
that was developed by ATO for research purposes. The file contains
just over 109,000 records of individual tax returns and detailed
information on income from various sources; different types of tax
deductions; taxable income; and the take-up of PHI by the
individuals. It also contains a limited number of demographic variables,
including gender, age group, and marital status.
In this paper, we focus on singles between 20 and 69 years old, who
were all subject to the same income threshold of \$50,000 for the
MLS. Therefore, the PHI take-up rate is expected to have a jump
around that level of the annual taxable income.
In the tax and transfer system or in the health
insurance premium regime in Australia, there is no other differential
treatment related to the PHI take-up. Other demographic covariates
(such as gender and age) would not generate discontinuity in the take-up
rate either as a function of the annual taxable income. So, in the
current PHI data, MLS seems to be the only factor responsible for the
jump in the take-up rate.

As a method of confidentialization, ATO ``perturbed'' the income
variables and the deductions, and provided the following information on the
way the data was perturbed: several random numbers within a specified
range for each individual were generated, which were converted into a
rate (equal probability of being positive or negative) and which was
then applied
to the various components of the tax return. These rates were applied
to the components in a way to try to maintain relationships with
similar items. This was achieved by grouping the components into three
broad categories: work or employment related income and deductions;
investment income and deductions; and business and other income
and deductions. Thus, there is some information about the measurement
errors in the income data, but the actual distribution of the measurement
errors is impossible to be identified based on the provided information.
The sample was further restricted to minimize the number of income
sources/deduction sources so that the distribution of the error term
could be more homogeneous, according to the following criteria:
(1)~Only those who had positive earnings as the only sources of income
were selected;
(2)~Individuals whose taxable income was not positive
(which means their total tax deductions were not less than their earnings)
were dropped; and
(3) We further dropped individuals whose nonwork related
deductions formed a significant part of their taxable income---specifically,
we dropped those individuals whose work related deductions were less
than 90
percent of earnings when the total deductions were more than 10 percent
of earnings; whose total deductions were over 50 percent of earnings; or
whose total deductions were all nonwork related and the total deductions
were over 10 percent of their earnings.

The final sample for analysis contains 9685 records of individual tax
returns. By a preliminary analysis, we found
that about $26\%$ of singles bought PHI in 2003--2004, and the PHI
take-up rates for those whose annual taxable incomes were below
\$50,000 and those whose annual taxable incomes were above
that level were quite different. The PHI take-up rate for the former group
was about $21\%$ and it was about $57\%$ for the latter group.
Because ATO perturbed the income data by multiplying each original
income observation by a random number, we used the income variable
in $\log$ scale in our analysis, so that the additive measurement
error assumption in (2) is valid here.

We then use our proposed jump detector (7) to estimate the jump
position, in which the possible jump location is searched within
$[10, 11.25]$ (or, equivalently, [\$22,026, \$76,879] of annual
taxable income). The results are shown in Figure~\ref{fig4}, where the
estimated function of $p(\cdot)$ by the local linear kernel (LLK)
smoothing procedure is shown by the solid line, the left-sided and
right-sided estimates of $p(\cdot)$ are shown by the dashed and dotted lines,
respectively, their difference is shown by the dot-dashed line at the
bottom of the plot, and the jump location estimate $\widehat{s}_n$
and the corresponding $95\%$ confidence interval for $s$ are shown
by the vertical dot-dash and long-dash lines, respectively. In the plot,
the related estimates look noisier near the right end because there
were fewer
people who had high incomes. The estimated jump location is
$\widehat{s}_n=10.9255$ ($\approx$\$55,575). The bandwidth
chosen by the bootstrap procedure is 0.0792. The $95\%$
confidence interval for $s$ computed by the proposed bootstrap
procedure is (10.8792, 11.0158) [$\approx$(\$53,061, \$60,827)],
which implies that the true jump location $s$ is significantly larger
than 10.8198 ($\approx$\$50,000). This finding confirms our
intuition that people usually act later than they are hit by the MLS.
The estimated jump magnitude by \eqref{eqjs} is 0.19. This number
shows that the local effect at the MLS tax policy discontinuity is quite
big. For individuals with only one income source, the policy can be
considered locally exogenous because the observations to the left and
right of
(but close to) the jump position are more or less homogeneous except
the policy treatment. It implies that, among the individuals whose
annual taxable income is around
\$55,575, MLS brings about an extra 19\% of them onto the private
health system. This
also implies a negative price elasticity of PHI demand
since the jump in the take-up rate can be seen as a response to a price
discount in the premium.

\section{Concluding remarks}
\label{secdiscussion}
We have proposed a generalized error-in-variab\-les jump regression
model for describing the relationship between people's annual taxable
income and the PHI take-up rate in Australia.
A novel jump detector is proposed as well, which can accommodate the
possible measurement errors. A~major feature of the proposed method
is that it does not require much prior knowledge on the measurement
error distribution, making it applicable in practice. Its
performance is evaluated by both numerical studies and
theoretical justifications.
By the proposed method, we found that the actual jump in the PHI
take-up rate, caused by the MLS tax policy, occurred at a larger
taxable income value than the threshold value used in the policy.

There is much room for further improvement of the current
method. First, the proposed jump detection method assumes that there is a
single jump point at an unknown location. By the framework of
jump regression analysis [cf. \citet{qiu2005}], it might be
possible to
extend it to cases when there are multiple jump points and the number
of jump points could be either known or unknown. Second, this paper
focuses on jump detection only. It requires much future research to
develop an
appropriate method to estimate a jump regression function from
observed data with measurement errors. Third, it might be important to extend
the current method to higher-dimensional cases.


\section*{Acknowledgments}
We thank the Editor, the Associate Editor, and
the two anonymous referees for many constructive comments and suggestions
which greatly improved the quality of the paper.

\begin{supplement}[id=suppA]
\stitle{Supplement to ``Jump detection in generalized error-in-variables regression with an application to Australian health tax policies''\\}
\slink[doi]{10.1214/15-AOAS814SUPP} 
\sdatatype{.pdf}
\sfilename{aoas814\_supp.pdf}
\sdescription{This supplemental file mainly gives the proof of Theorem~\ref{thm1}.}
\end{supplement}

%

\printaddresses
\end{document}